\documentclass{article}

\usepackage{arxiv}
\usepackage[utf8]{inputenc} 
\usepackage[T1]{fontenc}    
\usepackage{hyperref}       
\usepackage{url}            
\usepackage{booktabs}       
\usepackage{amsfonts}       
\usepackage{nicefrac}       
\usepackage{microtype}      
\usepackage{lipsum}
\usepackage{graphicx}
\usepackage{amsmath}
\usepackage{amssymb}
\graphicspath{ {./images/} }

\title{Confocal 3D reflectance imaging through multimode fibers without wavefront shaping}

\author{ 
    Szu-Yu Lee \\
    Harvard-MIT Health Sciences and Technology \\ 
    Massachusetts Institute of Technology \\
    Cambridge, MA 02139, USA \\
	\texttt{szuyul@mit.edu} \\
	\And
	Vicente J. Parot \\
	Institute for Biological and Medical Engineering \\
	Pontificia Universidad Católica de Chile \\
	Santiago 7820244, Chile \\
	\texttt{vparot@uc.cl} \\
	\AND
	Brett E. Bouma \\
	Institute for Medical Engineering and Science \\
	Massachusetts Institute of Technology \\
	Cambridge, MA 02139, USA \\
	\texttt{bbouma@mit.edu} \\
	\And
	Martin Villiger \\
	Wellman Center for Photomedicine \\
	Harvard Medical School and Massachusetts General Hospital \\
	Boston, MA 02114, USA \\
	\texttt{mvilliger@mgh.harvard.edu} \\
}

\hypersetup{
pdftitle={Confocal 3D reflectance imaging through multimode fibers without wavefront shaping},
pdfsubject={physics.optics},
pdfauthor={Szu-Yu Lee, Vicente J. Parot, Brett E. Bouma, Martin Villiger},
pdfkeywords={Multimode fiber, Transmission matrix, Confocal imaging, 3D reconstruction, Label-free imaging},
}

\begin{document}
\maketitle

\begin{abstract}
Imaging through optical multimode fibers (MMFs) has the potential to enable hair-thin endoscopes that reduce the invasiveness of imaging deep inside tissues and organs. Current approaches predominantly require active wavefront shaping and fluorescent labeling, which limits their use to preclinical applications and frustrates imaging speed. Here we present a computational approach to reconstruct depth-gated confocal images using a raster-scanned, focused input illumination. We demonstrate the compatibility of this approach with quantitative phase, dark-field, and polarimetric imaging. Computational imaging through MMF opens a new pathway for minimally invasive imaging in medical diagnosis and biological investigations. \\ 
\end{abstract}

\keywords{Multimode fiber \and Transmission matrix \and Confocal imaging \and 3D reconstruction \and Label-free imaging}

\section{Introduction}

The use of optical multimode fibers (MMFs) as imaging conduits could enable hair-thin minimally invasive endoscopes that provide visual access to otherwise difficult to reach tissue sites \cite{Keiser:14}. Imaging through MMF has shown many advantages over other optical endoscopes: pliable geometry and lower cost compared to wide-field rod-lens endoscopes, minimized probe size and variable sampling rate and working distance compared to micro-electromechanical-system- (MEMS) based endoscopes, and dense mode population over the core area compared to fiber bundles and multi-core fibers. Although current approaches to imaging through MMFs have small tolerance to bending and generally require a rigid shape after calibration \cite{Loterie:17}, progress with addressing this limitation \cite{PhysRevX.9.041050,matthes2020learning,li2020guidestar} highlights the promising biomedical significance of MMF imaging in neuroscience and imaging-guided surgery and biopsy \cite{PhysRevLett.109.203901,Turtaev2018,Vasquez-Lopez:18,Leite:21}.

Several strategies have been proposed to compensate the chaos caused by mode mixing and dispersion in MMF transmission, including wavefront shaping (WFS) \cite{Papadopoulos:12,Cizmar2012,Loterie:15,Ploschner2015,Gusachenko:17,Ohayon:18,Turtaev2018,Leite:21}, speckle imaging \cite{PhysRevLett.109.203901,Caravaca-Aguirre:19}, compressive imaging \cite{Amitonova2018,Choudhury2020}, and deep learning \cite{Borhani:18,Caramazza2019}. WFS, which is the primary strategy for endoscopic imaging of complex three-dimensional (3D) biological samples, generates a sharp distal focus through a MMF by taking into account the calibrated MMF transmission. In this approach, a distal focus is scanned through the sample for image formation. Combined with focusing through MMF using WFS, digital phase filtering on the detection pathway has been demonstrated to enhance the optical image contrast of non-fluorescent samples and allow optical sectioning to resolve axial structures \cite{Loterie:15} without the need for pulsed lasers as in time of flight methods \cite{Stellinga:19}. 

Unfortunately, WFS techniques require complex systems and encounter limitations in imaging speed, focus accuracy, and power efficiency. One fundamental limitation is the physical latency of wavefront shaping modules. While an unprecedentedly fast WFS technique has been introduced to enable real-time focusing through scattering media by using an ultra-high-speed 1D modulator \cite{Tzang2019}, imaging through MMF with this technique is associated with suboptimal focus contrast due to the uncompensated transverse dimension. Furthermore, most MMF imaging is so far geared towards preclinical imaging in mouse models and combines fluorescence labeling with physical scanning for high sensitivity and specificity. Imaging through MMFs for clinical diagnostic applications would benefit from exogenous-label-free imaging. Towards this end, Choi et al. reported on wide-field label-free imaging through MMF that avoids WFS and offers high imaging speed by employing galvanometer scanning mirrors and a turbid lens imaging algorithm \cite{PhysRevLett.109.203901}, but this approach does not provide optical sectioning or phase information, leading to low-contrast wide-field images.

This work builds on the simplicity of WFS-free MMF imaging and incorporates 3D optical sectioning to increase image contrast by introducing a new computational strategy that enables synthetic confocally gated imaging through MMF. The method achieves high-contrast reflectance imaging through point-spread function (PSF) engineering based on coherent transmission matrix (TM) operations. After initial calibration of a MMF – common to all current MMF imaging strategies – and obtaining its full forward TM by scanning a focal spot over the proximal side of the MMF, we measure the round-trip reflection matrix of the fiber including the sample at its distal end using the same proximal spot basis. Owing to reciprocity, the TM defines the bi-directional light transport through the MMF and describes light scrambling on both illumination and detection pathways \cite{Lee:20}. We then digitally compensate the effect of transmission through the MMF on the reflection matrix by left- and right-multiplication with the regularized matrix inversion of the TM and its transpose, respectively. The individually addressable recovered illumination and detection modes enable remote confocal gating of selected observation planes (OPs). Using numerical refocusing, a single measurement of the reflection matrix yields volumetric images of the sample, with field of view (FOV) and spatial resolution on an OP determined by the fiber numerical aperture (NA) and the solid angle subtended by the fiber core. Finally, we show imaging of complex samples with versatile optical contrasts including dark-field and birefringence imaging to demonstrate the potential for label-free endoscopic biomedical imaging. 

\section{Overview}
\label{sec:overview}
\begin{figure*}[t!] 
\centering
\includegraphics[width=6.5in]{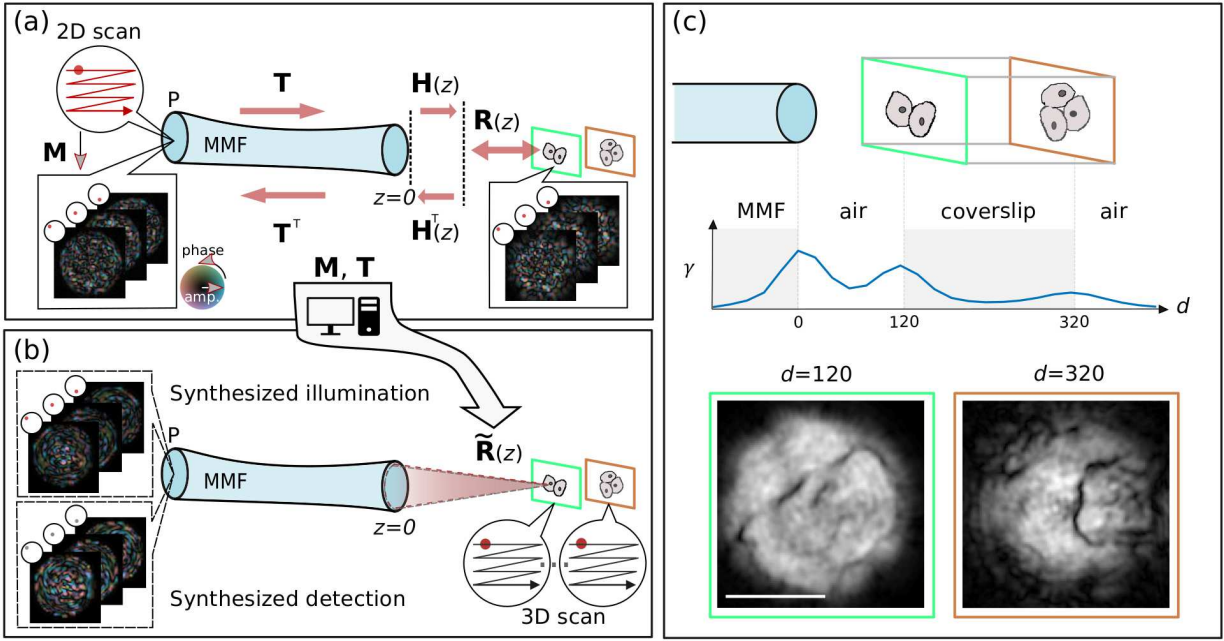}
\caption{Overview of computational confocal imaging through MMF. (a) We calibrate the MMF by measuring \textbf{T}. In the imaging phase, we measure the coherent round-trip \textbf{M} through the same set of proximal channels as used for calibrating \textbf{T}, but now in the presence of a distal object. The \textbf{H} matrix accounts for free-space propagation from the calibration plane to an OP. The sample is illuminated by different speckle realizations due to the proximal 2D scan, and \textbf{R} matrix denotes light-sample interaction. The measured \textbf{M} and \textbf{T} are used for computational reconstruction. (b) By modeling the round-trip light transmission with matrix multiplications, we can compensate the MMF scrambling using the measured \textbf{T} to isolate the reflection matrix $\tilde{\textrm{\textbf{R}}}$. The image of the sample is reconstructed from the diagonal of $|\tilde{\textrm{\textbf{R}}}|^2$, which is essentially synthesized confocal illumination and detection through each available channel. With numerical refocusing, we can complete 3D scan with a single measured \textbf{M}. The color map encodes complex values. amp.: amplitude. (c) Depth-gated confocal images and the corresponding $\gamma$ profile of a 3D-structured sample. The $d$ axis shows physical distance from the fiber facet. $d$ is in the unit of \textmu m. The scale bar is 50 \textmu m.}
\label{fig:overview}
\end{figure*}

The monochromatic optical transmission through a general medium from an input surface to an output surface can be expressed by a coherent matrix, specifying the amplitude and phase evolution of the transmitted field between pairs of input and output spatial channels. As illustrated in Fig. \ref{fig:overview}(a), we express forward light transmission through a MMF as matrix \textbf{T}. The measurement of \textbf{T}, by determining the amplitude and phase of the output speckle pattern arising from focal illumination at each independent transverse location on the input fiber facet, calibrates the transmission from the proximal (P) end to a distal calibration plane (at $z=0$). The MMF output speckle pattern on the calibration plane per each proximal input realization constitutes a column vector of \textbf{T}. In imaging mode with a sample at the fiber distal end, we illuminate and detect light from the fiber proximal end. This corresponds to a bi-directional light transport consisting of forward transmission through the MMF, free-space propagation to an OP modeled by Fresnel diffraction (\textbf{H}), speckle illumination on and backscattering from the sample, coupling back into the same MMF, and backward transmission to the proximal facet. The TM representing backward transmission $\textbf{T}^\textrm{T}$ is the transpose transformation of the forward TM $\textbf{T}$ due to reciprocity \cite{Lee:20}, and the overall round-trip reflection matrix \textbf{M}, describing optical transmission from and to the proximal side matrix formalism can be expressed as
\begin{equation}
\label{eqn:M}
\textrm{\textbf{M}} = \textrm{\textbf{T}}^\textrm{T} \textrm{\textbf{H}}^\textrm{T} \textrm{\textbf{R}} \textrm{\textbf{H}} \textrm{\textbf{T}},
\end{equation}
where \textbf{R} quantifies the backscattering process of the light-sample interaction in the spot basis on the distal OP. \textbf{R} has intrinsic transpose symmetry, $\textrm{\textbf{R}}=\textrm{\textbf{R}}^\textrm{T}$. Similar to \textbf{T}, each column of \textbf{M} is a proximally recorded speckle pattern per input realization. The transpose symmetry of \textbf{M} that follows from Eq.\ref{eqn:M} is experimentally attained as elaborated in the \nameref{sec:methods} section.

The computational reconstruction is illustrated in Fig. \ref{fig:overview}(b). With a previously measured \textbf{T}, we can digitally compensate the light scrambling during round trip MMF propagation and extract \textbf{R} from \textbf{M} by right- and left- multiplying it with the inverse and transpose inverse of \textbf{HT}, respectively. The \textbf{H} matrix can be numerically generated for arbitrary refocusing and controls the OP position (see Supplementary Materials S2). In practice, the measured \textbf{T} is generally non-square, ill-posed, and noisy, so we used Tikhonov regularization to approximate its inversion and transpose inversion, $\textrm{\textbf{T}}^\textrm{--1(tik)}$ and $\textrm{\textbf{T}}^\textrm{--T(tik)}$, with the regularization parameter set to $10\%$ of the largest singular value as justified by the L-curve method \cite{Hansen00thel-curve}. The approximated \textbf{R}, $\tilde{\textrm{\textbf{R}}}$, thus can be derived as
\begin{equation}
\label{eqn:R_extract}
\tilde{\textrm{\textbf{R}}} = \textrm{\textbf{H}}^\textrm{--T} \textrm{\textbf{T}}^\textrm{--T(tik)} \textrm{\textbf{M}} \textrm{\textbf{T}}^\textrm{--1(tik)} \textrm{\textbf{H}}^\textrm{--1} \approx \textrm{\textbf{R}},
\end{equation}
where matrices are defined regardless of basis representation. When the input and output of $\tilde{\textrm{\textbf{R}}}$ are both in the spot basis, an adequate high-contrast image reconstruction of the \textit{en face} scattering on the OP can be obtained by reshaping the intensity on the diagonal of $\tilde{\textrm{\textbf{R}}}$ into its corresponding 2D xy-layout. Physically, the diagonal elements of $\tilde{\textrm{\textbf{R}}}$ correspond to synthetic focal illumination and detection occurring through identical channels on the OP, creating a spatial confocal gating effect with a depth of focus determined by the effective NA available at each location on the OP. By varying \textbf{H}, we can numerically shift the OP along the optical axis to different distances and reconstruct the full addressable 3D image volume from a single measured \textbf{M}. 

\section{Methods}
\label{sec:methods}

\subsection{MMF calibration and sample reflectance measurement}
All experiments used a 1-m-long step-index MMF with $105$ \textmu m core diameter and a NA of 0.22 (FG105LCA, Thorlabs) that theoretically supports ${\sim550}$ guided modes per polarization at $\lambda = 1550$ nm. The fiber was coiled with a minimum radius of curvature of ${\sim50}$ mm. The monochromatic calibration matrix \textbf{T} was measured by sequentially probing the MMF input channels with a focal spot, while complex detection was used for all distal output channels concurrently. Each input and output channel included two orthogonal polarization states: horizontal (H) and vertical (V). The focal spot position on the proximal input side was indexed by $u$ and the speckle pattern exiting on the distal side was imaged with an off-axis holographic imaging system, whose object plane determined the calibration plane, approximately 100 \textmu m away from the MMF distal facet. Images were flattened into column vectors of \textbf{T} directly in the Fourier domain, with output channels indexed by $\nu_F$. The input and output spatial channels of \textbf{T} have been ordered first by spatial coordinate, then by polarization. 

In imaging experiments, we again sequentially coupled light into the MMF through the same set of proximal input states. We then recorded the round-trip light transmission on the proximal side by the same off-axis holography setup. To preserve the symmetry between the illumination and the detection configurations and to obtain a square matrix \textbf{M}, we sampled the recorded complex output fields at the ordered positions identical to the set of input states. Numerical corrections to compensate for the physical misalignment were then applied to the output channels of the measured \textbf{M} to accurately match the input channels and recover the underlying transpose symmetry as previously described \cite{Lee:20}. The detailed experimental setup and data processing are in Supplementary Materials S1.

\subsection{Confocal image reconstruction}
Using the sample measurement \textbf{M} and the pre-measured \textbf{T}, we then computed $\tilde{\textrm{\textbf{R}}}$ following Eq. \ref{eqn:R_extract}. The free-space propagation matrix \textbf{H}($z$) from the calibration plane to a selected OP($z$) was defined as a diagonal matrix in $\nu_F$ (illustration in Fig. S1(b-1) and details in Supplementary Materials S2), considering the medium refractive index. The input and output basis of $\tilde{\textrm{\textbf{R}}}$ were converted from $\nu_F$ to $\nu$ by multiplication with a pre-computed inverse discrete Fourier transform matrix. A 2D confocal intensity image \textbf{I} of sample reflectance at the OP was reconstructed by reshaping of the diagonal of $\tilde{\textrm{\textbf{R}}}$ as 
\begin{equation}
\label{eqn:bright-field confocal from R}
\textrm{\textbf{I}}(x,y) = |\tilde{\textrm{\textbf{R}}}[\nu(x,y),\nu(x,y)]|^2,
\end{equation}
where the point $(x,y)$ is mapped from the distal spatial channel $\nu$ to real-space coordinates, and $[\cdot]$ indicates matrix entries, arranged in row and column. For polarization-preserving samples, reconstructed images of co-polarized illumination and detection channels are identical and incoherently summed to increase signal. This computation was repeated for multiple values of $z$ to generate 3D images from a single reflectance measurement \textbf{M} with depth expressed in $d$ (referenced to distal facet). Intensity images were converted to base-10 logarithmic scale for display.

\subsection{Wide-field image reconstruction}
From the same measured \textbf{M}, we can also obtain wide-field imaging that is equivalent to the turbid lens imaging algorithm \cite{PhysRevLett.107.023902}. We compensated for the reverse MMF transmission of reflectance from the sample under the variety of speckle illuminations, and then incoherently averaged the reflectance to statistically compose a uniform illumination. In terms of matrix operations, we left-multiply Eq. \ref{eqn:M} with $\textrm{\textbf{H}}^\textrm{--T}\textrm{\textbf{T}}^\textrm{--T(tik)}$ 
\begin{equation}
\label{eqn:wide-field matrix}
\tilde{\textrm{\textbf{R}}} \textrm{\textbf{H}} \textrm{\textbf{T}}
= \textrm{\textbf{H}}^\textrm{--T}\textrm{\textbf{T}}^\textrm{--T(tik)} \textrm{\textbf{M}},
\end{equation}
where each column of the matrix product is the sample reflection resulting from a distinct speckle illumination. Wide-field images were reconstructed by integrating the absolute square of $\tilde{\textrm{\textbf{R}}}$\textbf{HT} along the input dimension into a single column vector
\begin{equation}
\label{eqn:wide-field imaging with RHT}
\sum_{u} |\tilde{\textrm{\textbf{R}}} \textrm{\textbf{H}} \textrm{\textbf{T}}(:,u)|^2,
\end{equation}
and reshaping the vector back to 2D coordinates. \textbf{T}(:,$u$) means $u^{th}$ column vector of \textbf{T}. To simplify computation, the matrix product \textbf{HT} in Eq. \ref{eqn:wide-field matrix} was assumed to be unitary, so that by Parseval’s theorem the integrated row intensity of $\tilde{\textrm{\textbf{R}}}$\textbf{HT} is identical to that of $\tilde{\textrm{\textbf{R}}}$. Confocal and wide-field images from the same $\tilde{\textrm{\textbf{R}}}$ can be thereafter fairly compared.

\subsection{Quantitative phase imaging}
In confocal imaging, due to the complex nature of $\tilde{\textrm{\textbf{R}}}$, quantitative phase imaging of thin samples can be accomplished by taking the complex values of the diagonal elements of a computed $\tilde{\textrm{\textbf{R}}}$ to form a complex 2D image (\textbf{X}),
\begin{equation}
\label{eqn:phase-contrast confocal from R}
\textrm{\textbf{X}}(x,y) = \tilde{\textrm{\textbf{R}}}[\nu(x,y),\nu(x,y)],
\end{equation} 
where the amplitude encodes the absolute reflectivity, and the phase quantifies changes in the wavefront of light propagating through the specimen and back. 

\subsection{Dark-field imaging}
Off-diagonal elements of $\tilde{\textrm{\textbf{R}}}$ also contain abundant information of sample optical properties, which can be extracted through manipulations on $\tilde{\textrm{\textbf{R}}}$. For instance, each column of $\tilde{\textrm{\textbf{R}}}$ represents the scattering at OP in response to an illumination focused on a single channel $q$. Instead of collecting the intensity at the corresponding location on the matrix diagonal, the intensity of surrounding output channels $\nu$ was summed with weights $L(\nu,q)$ given by their Euclidean distance from the input channel on the xy-plane (the on-diagonal confocal signal thus has a zero weight) with a cutoff radius of 12 \textmu m, and normalized by the overall intensity 
\begin{eqnarray}
\label{eqn:dark-field confocal from R}
&& \textrm{\textbf{S}}(x,y) = \frac{\sum_\nu L(\nu(\zeta,\xi),q(x,y)) \times|\tilde{\textrm{\textbf{R}}}[\nu(\zeta,\xi),q(x,y)]|^2}{\sum_\nu |\tilde{\textrm{\textbf{R}}}[\nu(\zeta,\xi),q(x,y)]|^2}, \nonumber \\ 
&& L(\nu(\zeta,\xi),q(x,y)) \equiv \sqrt{(\zeta-x)^2 + (\xi-y)^2}
\end{eqnarray}
where the Cartesian point $(\zeta,\xi)$ maps to the distal channel indexed at $\nu$. We named this metric scattering contrast (\textbf{S}). Since $\tilde{\textrm{\textbf{R}}}$ is transpose-symmetric, interchanging the illumination and detection renders identically reconstructed images. To compare with microscopy, for each location in the image plane, the scattering contrast \textbf{S} is essentially the combination of focused illumination and ring-shaped detection PSF, which captures positive signals from the boundaries of sample heterogeneity and is analogous to a dark-field confocal imaging scheme \cite{Scoles:13}. 

\subsection{Polarization contrast}
So far, computation of images from $\tilde{\textrm{\textbf{R}}}$ only considered co-polarized illumination and detection, where each distal spatial channel $\nu$ degenerates into $\nu_H$ and $\nu_V$, and entries corresponding to input and output channels were used with the same polarization state. For birefringent samples such as collagen, illumination through a channel in a certain polarization state may induce cross-polarized backscattering. While $\tilde{\textrm{\textbf{R}}}$ is symmetric and has inputs and outputs ordered first by coordinates and then by polarization, the diagonals of the two off-diagonal matrix quadrants represent cross-polarized detection, and the sample birefringence at individual image positions $(x,y)$ can be resolved and characterized by assembling 2-by-2 Jones matrices,
\begin{equation}
\label{eqn:Jones matrices from R}
\textrm{\textbf{J}}(x,y) = 
\begin{bmatrix}
\textrm{\textbf{J}}_{11} & \textrm{\textbf{J}}_{12}\\
\textrm{\textbf{J}}_{21} & \textrm{\textbf{J}}_{22}\\
\end{bmatrix}
=
\begin{bmatrix}
\tilde{\textrm{\textbf{R}}}[\nu_H,\nu_H] & \tilde{\textrm{\textbf{R}}}[\nu_H,\nu_V]\\
\tilde{\textrm{\textbf{R}}}[\nu_V,\nu_H] & \tilde{\textrm{\textbf{R}}}[\nu_V,\nu_V]\\
\end{bmatrix},
\end{equation}
in the basis of orthogonal linear polarization states. From the Jones matrix at each channel, a retardation matrix was isolated using polar decomposition. Owing to the intrinsic transpose symmetry, the resulting matrix describes a linear retarder that can be characterized by its amount of retardance (ret) $\delta$ and optic axis (OA) $\phi$ orientation. Endogenous contrast within birefringent samples can thus be retrieved from this polarization-diverse measurement. 

\section{Results}

\subsection{Depth-gated imaging through MMF without WFS}
\label{sec:confocal vs. wide-field comparison}
To demonstrate the depth gating effect of our computational reconstruction, we imaged a 3D-structured sample through the MMF, as shown in Fig. \ref{fig:overview}(c). The sample is a coverslip in air with buccal epithelial cells deposited on both surfaces. We computed the confocal image for each OP at varying distance $d$ from the MMF distal facet ($d=0$) and calculated the corresponding integrated reflectivity ($\gamma$) by summing the intensity over the entire \textit{en face} image. The $\gamma$ versus depth profile reveals three separated peaks, which inform reflective MMF facet and coverslip surfaces, and their axial positions at $d=0$, $120$, and $320$ \textmu m considering medium refractive index. The green and brown insets show high-contrast images of cells on the front ($d=120$ \textmu m) and back ($d=320$ \textmu m) surfaces of the coverslip, respectively. Our matrix approach, which combines confocal gating with numerical refocusing, thus enables 3D scan from a single measured \textbf{M} without WFS. For more details and additional results on this experiment, please see Supplementary Materials S6.

\begin{figure}[t!] 
\centering
\includegraphics[width=3.5in]{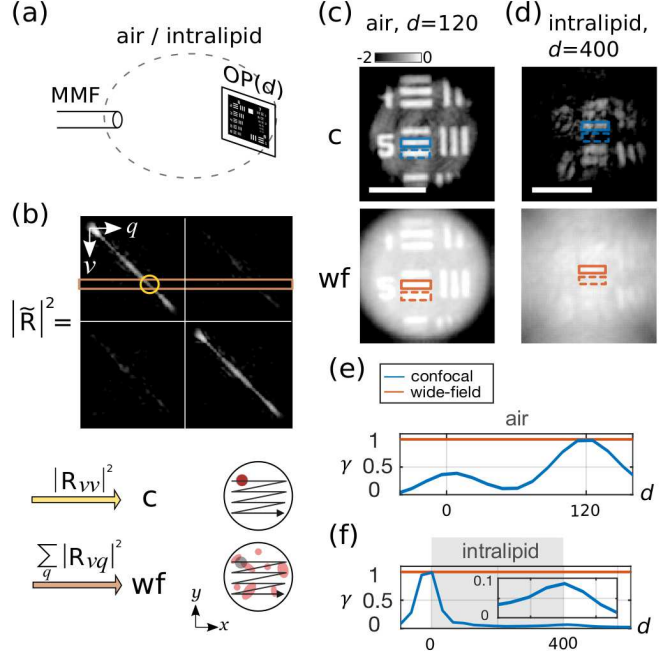}
\caption{Depth gating of computational confocal imaging. (a) Imaging geometry: a resolution chart was imaged at a distance in front of the distal MMF facet. (b) From the computed sample reflectivity matrix $\tilde{\textrm{\textbf{R}}}$, pixels of a confocal image were obtained by taking the intensity of diagonal elements, and of a wide-field image by taking the energy of row vectors. Images at multiple depths were computed from the same measured \textbf{M} with numerical refocusing. The image diagrams show the fundamental difference in illumination (red) and detection (gray) patterns of computational image formation. c: confocal; wf: wide-field. (c)-(d) Confocal images showed high, axially localized contrast compared to wide-field images, both in (c) air (at $d=120$) and in (d) intralipid media (at $d=400$). Images show logarithm of normalized intensity. The solid line (or dashed line) boxes indicate chrome pattern (or glass substrate) areas for image contrast quantification. (e)-(f) The normalized integrated reflectivity plots of confocal (blue) and wide-field (orange) images in (e) air and in (f) intralipid media. $d$ is in the unit of \textmu m. The scale bars are $50$ \textmu m.} 
\label{fig:confocal_widefield}
\end{figure}

To further evaluate the confocal gating effect, we imaged a USAF resolution chart (R1D21P, Thorlabs) in different media and distances $d$ through the MMF, as sketched in Fig. \ref{fig:confocal_widefield}(a). In each medium, a sample reflectance matrix $\tilde{\textbf{R}}$ was computed from a single measured \textbf{M} for each OP at varying depth to find in-focus position, and processed to reconstruct confocal and wide-field images for direct comparison, as shown in Fig. \ref{fig:confocal_widefield}(b). The pixel-wise illumination and detection are also illustrated. Since the chart has a binary reflectance pattern across its surface, we can quantify the intensity image contrast as
\begin{equation}
\label{eqn:quantify image contrast}
\varsigma = \frac{I_p - I_g}{I_p + I_g},
\end{equation}
where $I_p$ and $I_g$ are the intensities of the chrome pattern and the glass substrate, respectively. In Fig. \ref{fig:confocal_widefield}(c) and (d), $I_p$ and $I_g$ are averaged within selected regions of interest indicated by solid-line and dashed-line boxes for the chrome and glass substrate areas, respectively. In each imaging condition and modality, we calculated the corresponding $\gamma$ profile, which is normalized by the highest value along the axial OP positions. 

To directly study the confocal gating efficiency, we imaged the chart placed at $d=120$ \textmu m in air. In Fig. \ref{fig:confocal_widefield}(c), the confocal method renders the chart patterns with high contrast due to the rejection of background signals from reflection at the MMF facet. The confocal image achieved a contrast of 0.9, close to the expected value of ${\sim}0.92$, assuming full reflection from the chrome pattern and 4\% reflection at the air-glass interface. In Fig. \ref{fig:confocal_widefield}(e), the profile reveals two prominent and separated peaks at $d=0$ and $d=120$ \textmu m, corresponding to the MMF facet and the resolution chart, respectively. This shows the optical sectioning of confocal gating.

To test the capacity of computational confocal gating in the presence of additional sample scattering, we imaged the chart placed at $d=400$ \textmu m in agarose gel mixed with $\textrm{2\:wt.}\% $ intralipid, which corresponds to ${\sim}0.36$ mean free paths. In Fig. \ref{fig:confocal_widefield}(d), the reduced confocal image quality may be due to: intralipid scattering that distorted the wavefront and the reconstructed images, degraded spatial resolution upon beam divergence at large $d$ (elaborated in the following section), or the limitation of numerical \textbf{H} when the OP is far from the calibration plane. Despite the medium scattering and lower signal to background ratio when the chart is far from the facet, the confocal image maintained a high contrast of 0.96, compared to a theoretical value of ${\sim}0.99$ (assuming $0.4\%$ reflection at the gel-glass interface). This could be attributed to the immersion gel having a refractive index closer to the glass material of chart and MMF, and the specular reflection from the chart substrate and MMF facet is attenuated. Meanwhile, the chart patterns still have full reflection, thereby resulting in a better image contrast. In Fig. \ref{fig:confocal_widefield}(f), the peak in the $\gamma$ profile at $d=400$ \textmu m corresponds to the chart, and, although weak, precisely informs on its physical location when assuming the medium refractive index to be 1.4. These results evidence the effective suppression of out-of-focus scattering and reflection without active WFS.

To theoretically compare the confocal method to the wide-field processing, one can juxtapose Eqs. \ref{eqn:wide-field matrix} and \ref{eqn:R_extract} to find that the matrix multiplication also on the right side of \textbf{M} pre-compensates the light scrambling effect of the MMF forward transmission, and synthesizes sharp foci through the MMF on a selected OP. In contrast, the wide-field processing of $\tilde{\textrm{\textbf{R}}}$ corresponds to speckle illumination, as illustrated in Fig. \ref{fig:confocal_widefield}(b). When imaging in air as in Fig. \ref{fig:confocal_widefield}(c), while the pattern with wide-field imaging stands out from the background on the OP at $d=120$ \textmu m, the strong background results in a low contrast of 0.48, and the corresponding $\gamma$ profile in Fig. \ref{fig:confocal_widefield}(e) remains a constant throughout the entire observation range. On the other hand, the wide-field image in Fig. \ref{fig:confocal_widefield}(d) can barely distinguish the pattern from the background, resulting in poor contrast of 0.11. The uniform $\gamma$ plot of wide-field imaging in Fig. \ref{fig:confocal_widefield}(f) again exposes the lack of optical sectioning. 

\subsection{Efficient imaging with flexible reconstruction, field of view, and spatial resolution}
\label{sec:FOV and resolution}
Imaging through a MMF using WFS typically scans a focus along a pre-defined scanning trace and records a single image point from each focus location. In contrast, our method illuminates the sample with a sequence of MMF-induced speckle patterns and utilizes the camera for parallel sampling of all addressable locations in the imaging volume. This allows arbitrary definition of sampling grid and working distance in post processing of single measurement of \textbf{M} (elaborated in Supplementary Materials S3). As light diverges upon exiting the MMF distal end governed by the fiber NA, computational reconstruction can adapt to a growing FOV with increasing OP distance from the fiber facet. Here, we demonstrate this flexible reconstruction in MMF reflectance imaging and evaluate the resulting FOV and 3D spatial resolution as a function of distance to the tip of the fiber.   

\begin{figure}[t!] 
\centering
\includegraphics[width=3.5in]{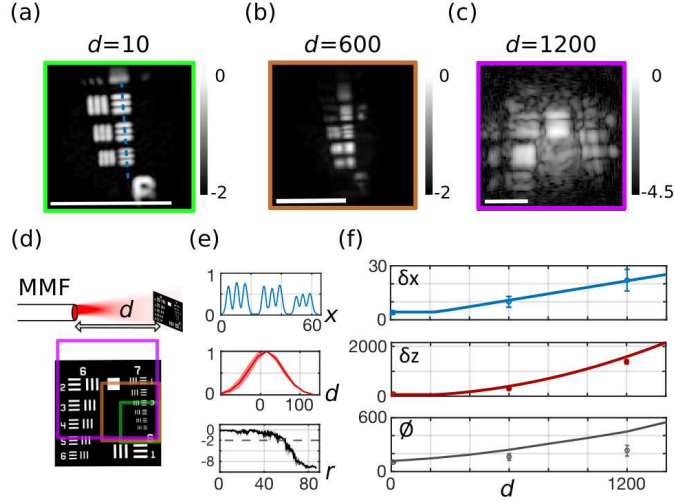}
\caption{Characterization of FOV and resolution of computational MMF imaging. (a)-(c) In-focus confocal intensity images of an USAF target at a distance (a) $d=10$ (b) $d=600$ (c) $d=1200$ \textmu m away from fiber facet. The green, brown, and purple boxes correspond to the ones in (d) that highlight the areas of the target. At $d=1200$ \textmu m, the speckle background reduces the image quality, and the dynamic range of the color bar is increased to reveal weak signals. (e) Analysis of imaging performance at $d=10$ \textmu m distance. The blue plot shows the normalized intensity profile in linear scale along the dashed line in the image in (a); the red plot indicates the $\gamma$ at varying $d$, and its FWHM estimates the axial resolution, with the shadowed areas indicating the standard deviations around the mean value of several independent realizations; the black curve shows the radial mean intensity in logarithmic scale, with 1\% cut-off at dashed line roughly equal to the fiber radius. (f) Experimental 3D resolution and FOV at various $d$ and corresponding theoretical values. All spatial dimensions are in the unit of \textmu m. The scale bars are 100 \textmu m.}
\label{fig:resolution_chart}
\end{figure}

To mimic endoscopic imaging with a variable working distance, a resolution chart was mounted on a translation stage and positioned at different distances $d=10$, $600$, or $1200$ \textmu m. For each $d$, a round-trip \textbf{M} was measured and computational reconstruction with numerical refocusing was utilized to locate the axial position of the resolution chart. Figure \ref{fig:resolution_chart}(a)-(c) show the in-focus confocal images of different chart areas (color boxes in (d)) with physical dimensions of $123$, $184$, and $247$ \textmu m, when $d=10, 600$, and $1200$ \textmu m, respectively. The illumination power on the chart was kept at ${\sim}0.5$ mW. Due to the beam divergence and limited laser power, camera exposure time was increased from $200$ \textmu s up to $1$ ms for larger $d$ to compensate for the declining photon collection. We filled the space between the fiber and the chart with index-matching gel (G608N3, Thorlabs) to mitigate the specular reflection from the MMF distal facet. In Fig. \ref{fig:resolution_chart}(a)-(c), imaging from farther away captures a more complete picture, as the FOV expands with increasing $d$. However, this comes at the expense of spatial resolution and collected reflectance power, as the patterns are severely blurred at $d=1200$ \textmu m, and background speckle becomes apparent. The finest detail of the chart, element 6 in group 7, can be resolved when the MMF is in close proximity of the facet, $d=10$ \textmu m, where the FOV is determined on a lower bound by the fiber core size. 

To quantify the spatial resolution at varying $d$, we inferred the lateral resolution, $\delta x$, from the smallest resolvable pattern on the chart, as shown by example dashed blue line and its linear intensity profile plot in \ref{fig:resolution_chart}(e). Also, since the chart serves as a sharp edge in the axial direction, we utilized the FWHM of the $\gamma$ profile around the focused $d$ to measure the axial resolution $\delta z$. The FOV of each computed image was characterized by its diameter $\varnothing$, set as twice the radius where the radially averaged intensity dropped below 1 \% of the center. We conducted several imaging realizations and computed corresponding axial profile and radial mean intensity at $d=10$ \textmu m for statistical analysis, as plotted in red and black curves in \ref{fig:resolution_chart}(e), respectively. For comparison, the theoretically expected spatial resolutions were derived considering the effective on-axis NA defined by the minimum between the fiber NA and the solid angle subtended by the fiber core at the corresponding distance from the facet. The theoretical FOV was estimated using $\textbf{T}$ by measuring the radial extent of averaged synthetic illumination patterns away from the facet (further details on theoretical resolution and FOV are presented in Supplementary Materials S4). The overall quantification results are shown in Fig. \ref{fig:resolution_chart}(f). While the experimental spatial resolutions are consistent with diffraction-limited theoretical values, the experimentally determined FOVs are up to 50 \% smaller than expected with increasing distance. This may be due to the low light collection efficiency of reflectance from distant objects or the limitation of numerical \textbf{H}. Nevertheless, the agreement in scaling properties of experimental and theoretical values corroborates the flexibility in addressable spatial dimensions given by the degrees of freedom guided through the MMF. These results demonstrate the convenience of reconstructing the entire sample volume without a pre-defined scan pattern in practical setting where the sample distance is unknown. Moreover, this flexibility of the matrix approach allows confocal image reconstruction from partial measurements of \textbf{M} with illumination through only a subset of proximal spatial channels (as shown in Supplementary Materials S5). While reconstruction from partial measurements compromises background suppression, it accelerates the volume rate, which may be critical for real-time applications.

\subsection{Multi-modal computational confocal imaging of label-free complex samples}
\label{sec:complex sample imaging}
To improve specificity in confocal MMF imaging without WFS for visualizing unlabeled biological specimens in a reflection mode, we leveraged the matrix approach to generate diverse contrasts from a measured round-trip \textbf{M} by applying different post-processing, and synthesized multiple imaging modalities to create signal specificity. Different strategies for image formation are described in the \nameref{sec:methods} section and illustrated in the following figures. The multi-contrast images in addition to the intensity images of the 3D imaging example in Fig. \ref{fig:overview}(c) can be found in the Supplementary Materials S6.

\begin{figure*}[t!] 
\centering
\includegraphics[width=6.5in]{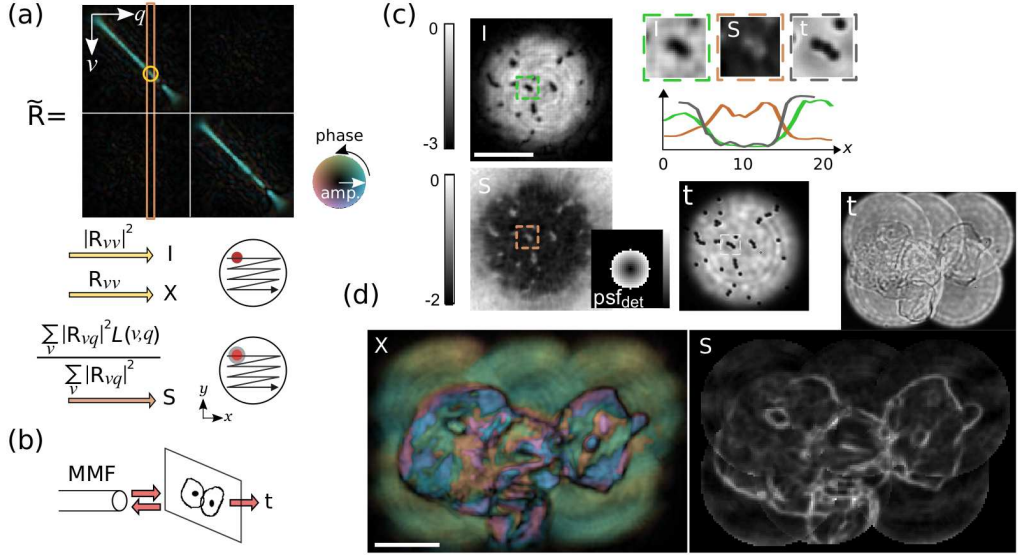}
\caption{Multi-modal MMF imaging of unlabeled samples including confocal intensity, quantitative phase imaging, and dark-field scattering imaging. (a) The diagonal elements of $\tilde{\textrm{\textbf{R}}}$ and their intensity correspond to pixels in confocal intensity (\textbf{I}) and complex (\textbf{X}) images, respectively. PSF engineering by an appropriate weighting function applied using Eq. \ref{eqn:dark-field confocal from R}, generates scattering contrast (\textbf{S}). (b) Sample arrangement at the MMF distal end. Transmission images (t) serve as ground truth for verification. (c) Confocal intensity and scattering contrast images of $3$ \textmu m polystyrene beads. The ring-shaped detection PSF is shown in the inset. Imaging based on scattering contrast features slightly better resolving power, judging from the plotted line profiles of beads in zoom-in dash-lined boxes of different modalities. (d) Stitched images of buccal epithelial cells with phase and dark-field contrasts. (left) Phase contrast depicts nuclei and intracellular morphology. (right) Dark-field scattering contrast reveals positive signals at cell boundaries and membrane roughness. The scale bars are $50$ \textmu m.}
\label{fig:bead_cell_imaging}
\end{figure*}

We started with non-birefringent samples to demonstrate multi-contrast imaging. Figure \ref{fig:bead_cell_imaging} (a) shows a typical reflection matrix $\tilde{\textrm{\textbf{R}}}$. The sample arrangement for these experiments, shown in Figure \ref{fig:bead_cell_imaging} (b), allowed imaging of a sample on a microscope glass slide in a reflection mode through the MMF, and also in transmission mode (t) with the distal imaging system and bright-field illumination through the MMF as ground-truth images. Figure \ref{fig:bead_cell_imaging} (c) shows the images of a monolayer of $3$ \textmu m polystyrene beads spread on the surface of a microscope slide and imaged in air at $d=100$ \textmu m. Since the reflectivity of the beads is orders of magnitude lower than that of the air-glass interface, the obtained round-trip $\tilde{\textrm{\textbf{R}}}$ on in-focus OP has diagonal elements dominated by the specular reflection from the glass slide, resulting in beads silhouetted against the glass signal in the confocal intensity image (\textbf{I}) and featuring negative contrast, similar to other reports of reflectance imaging through optical fibers \cite{PhysRevLett.107.023902,Loterie:15,Yoon2017}. With the full knowledge of $\tilde{\textrm{\textbf{R}}}$ and following Eq. \ref{eqn:dark-field confocal from R}, we are able to extract scattering signal specifically from the beads and create a dark-field image (\textbf{S}) through numerical engineering of the system detection PSF (sketched in the inset). Physically, bead-induced roughness distorted the specular reflection wavefront, which made returning light partially cross-coupled to neighboring spatial channels. Intriguingly, the cross-coupling signals that delineate the beads provided a slightly higher resolving power than the confocal intensity image, as verified by comparing the line profiles of clustered particles. This exemplifies the benefit of computational reconstruction, whereas physical implementation of dark-field imaging would traditionally require an annular filter, axicon lens, or customized pinhole, and increase the system complexity \cite{Cizmar2012, Scoles:13, Liu2020}.  

To demonstrate multimodal MMF imaging including phase and dark-field imaging from the same measurement of an unlabeled biological specimen, human buccal epithelial cells were smeared on microscope glass slide and placed at $d=120$ \textmu m in air. The sample was laterally translated to image several overlapping areas, and at each lateral location a round-trip \textbf{M} was measured to reconstruct the corresponding image. Multiple images were then stitched together to make a composite image with a wider FOV. Fig. \ref{fig:bead_cell_imaging} (d) shows phase contrast (left, \textbf{X}), revealing the contour of cellular membranes and nuclei in its amplitude (coded in brightness), likely because they deflect the focused illumination which attenuates the reflected signals, thereby resulting in negative contrast. Furthermore, as shown by its color-coded phase, the variation in sample thickness or refractive index inhomogeneity provides an intrinsic phase contrast of the unlabeled sample likely caused by sub-cellular structures, with information not contained in the intensity image alone. The scattering contrast image (right, \textbf{S}) delivers complementary information by positively outlining the cellular membrane morphology along with some cytoplasmic organelles that can be roughly correlated with the transmission image. 

\begin{figure}[t!] 
\centering
\includegraphics[width=3.5in]{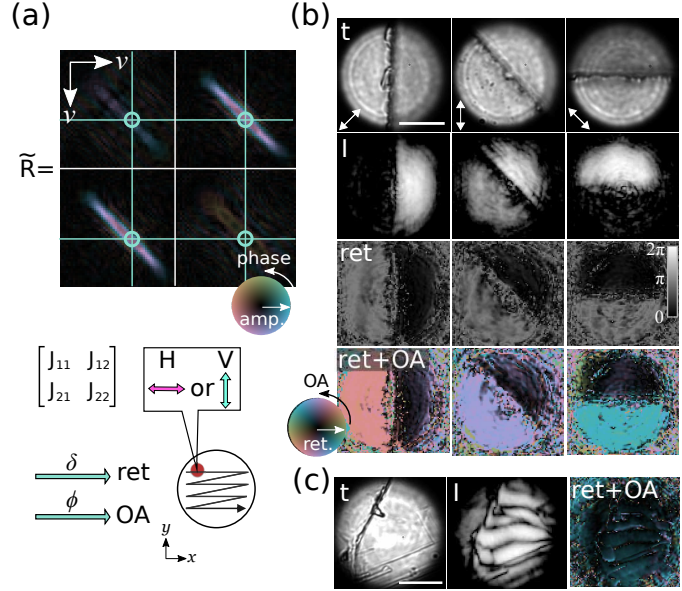}
\caption{Birefringence imaging of anisotropic samples through the MMF by utilizing reflection matrices. (a) Illustration of processing the matrix for computational polarization sensitivity. The full reflection matrix $\tilde{\textrm{\textbf{R}}}$ on an OP provides both co- and cross-polarization entries for assembling a Jones matrix at each spatial channel, which informs on retardation and OA orientation. (b) Accuracy evaluation of anisotropy reconstruction using a QWP with slow axis (white arrows) oriented at 45\textdegree, 90\textdegree, and 135\textdegree\ with respect to the x axis. While the confocal intensity images only show the sharp edge of the plate, the retardation and OA images unveil its intrinsic optical properties. (c) Birefringence imaging through the MMF of cholesterol crystal precipitated on a microscope slide. The visible fringes are attributed to thin film interference of the crystal and glass surfaces. The scale bars are 50 \textmu m.}
\label{fig:PS_imaging}
\end{figure}

To demonstrate polarization sensitive (PS) computational imaging through MMF based on our matrix approach and Eqs. \ref{eqn:Jones matrices from R}, as illustrated in Fig. \ref{fig:PS_imaging}(a), we obtained reflection matrices of anisotropic materials including quarter-wave plate (QWP) and cholesterol crystals through the MMF. To validate quantitative retardation and OA measurements, a QWP (WPQ501, Thorlabs) was placed on a microscope slide and its proximal reflection was measured. As show in Fig. \ref{fig:PS_imaging}(b), the edge of the wave plate was imaged in different orientations to verify the OA orientation retrieved from polarization analysis. One \textbf{M} was measured in each orientation. Due to the round-trip light propagation, the QWP has an effective half-wave retardation, which leads to full attenuation in the co-polarized detection for confocal intensity images when the slow axis is 45\textdegree\,to the H or V polarizations, and partial attenuation in between. Consistently, the corresponding retardance images reveal a constant $\pi$ rad retardation of the wave plate regardless of the orientation. On the other hand, the color-coded OA images (combined with brightness-coded ret images) show a rotating OA of the QWP with orientation exactly the same as the slow axis angle. Note that the OA colormap has a periodicity of $\pi$ instead of $2\pi$ used in phase colormaps. Figure \ref{fig:PS_imaging}(c) shows another example with home-made plate-like cholesterol crystals (S25677, Fisher Science Education) on a microscope slide, which has a much weaker retardance due to its small thickness (tens of \textmu m) yet uniform optic axis orientation. Judging from the values of retardation, the crystal may be thicker towards the bottom of the image. Since the crystal thickness is much smaller than the confocal gate, interference between the front and back surfaces results in \textit{en face} fringes. A tighter confocal gate may be achieved by switching to MMFs with higher NA or choosing a shorter operating wavelength.

\section{Discussion}
Computational confocal imaging through MMFs is a novel matrix-based method to obtain depth-gated images using a proximal scanning spot basis for reflectance measurement without WFS, yielding multimodal 3D reflectance of unlabeled samples including confocal intensity, quantitative phase, dark-field, birefringent retardance, and optic axis contrast modalities. Despite the many documented imaging through MMFs, this is the first report of numerical PSF engineering, phase, and PS imaging through MMF in a reflection geometry. 

High-contrast imaging through MMF frequently relies on fluorescent labeling or is operated in a transmission regime \cite{Amitonova2018,Morales-Delgado:15,Ohayon:18,Turtaev2018,Ploschner2015}, which is incompatible with practical endoscopic applications. Fluorescence scanning microendoscopy furthermore suffers from photobleaching \cite{OH2013760, Gu2014}. Our computational imaging approach instead efficiently extracts weak elastic scattering signals from unstained samples and operates in a reflection regime, making it  favorable to practical endoscopic applications. The matrix approach moreover offers an elegant way of achieving full polarization management and leverages polarization as additional contrast mechanism. In comparison, WFS for physical focusing through MMF typically addresses only a single polarization state \cite{Loterie:15,Turtaev2018}, to avoid the hardware complexity required for full polarization-control \cite{Ploschner2015}. While PSF engineering through complex media in a transmission regime has been documented \cite{Boniface:17}, our method here avoids SLM/DMD and optimization. More broadly speaking, illumination and detection with any respective PSF and in any polarization state can be readily engineered by weighting the entries of $\tilde{\textrm{\textbf{R}}}$ accordingly. 

The matrix approach employs a simple proximal spot basis for illumination, which relaxes hardware requirements by accepting any 2D scanning module and avoids the need for WFS. The limiting factor in imaging speed of this work is the InGaAs-camera frame rate of 120 Hz, which may be directly improved by an order of magnitude by replacing it with a faster one 
or by shifting the operation wavelength towards visible wavelengths with more and even faster camera options. Recently, MMF calibration covering 256 degrees of freedom within only 34 ms has been demonstrated by using a field programmable gate array (FPGA) to address the general latency issue in hardware communication \cite{Caravaca-Aguirre:13}, which exemplifies the efficacy of program optimization and is readily applicable to speed up our implementation. Our approach also allows image formation from a partial round-trip measurement with as few as 200 input realizations, which trades off slight image quality for ${\sim85}\%$ measurement time reduction. Because the reconstruction of individual spatial channels is independent from other channels, this offers high potential for parallelization of the processing using GPU acceleration. With careful engineering of the data acquisition and processing pipeline, fast video-rate imaging should be achievable. Fundamentally, imaging speed of our MMF imaging method is limited by computational complexity, and no longer by hardware as for WFS methods.

MMF imaging has a notorious intolerance to small fiber perturbations such as bending or looping. Even small fiber alterations typically require MMF re-calibration, which is exceedingly difficult in an endoscopic setting \cite{Lee:20}. This admittedly remains the most fundamental limitation towards practical use of flexible MMF endoscopy. In our experiments, the calibrated 1-m-long MMF was fixed looping on the optic table and remained stable for several hours without perceivable TM change such that the same measured TM could be used for imaging different samples through the MMF. In a practical setting, an ultra-thin MMF could be mechanically shielded inside a rigid needle or hypodermic tubing to enable high-quality imaging through MMF without re-calibration. Furthermore, several promising strategies are being pursued to address the need for TM calibration without physical access to the distal fiber end: the installation of carefully designed passive optics or guide star to the MMF distal tip \cite{ Gu:15, PhysRevX.9.041050, li2020guidestar}, compressive sampling of TM with sparsity constraint \cite{Li2021}, or the use of graded-index MMF which has increased robustness of light transport to bending deformations.

The disclosed method offers 3D confocal imaging through MMFs with high signal specificity, yet is less hardware-demanding than common WFS methods. This may expedite or create applications of minimally invasive MMF-based endoscopy in biomedicine, where probe size and cost are critical factors. For instance, deep brain imaging in neurosurgery, on-site inspection in needle biopsy, collagen imaging in arthroscopy, and tympanic cavity imaging in middle and inner ear surgery are potential uses of the technique. The same methodology may also be extended to optical imaging through other complex or turbid media, or other imaging technologies such as ultrasound tomography.

\section{Conclusion}
Accurate knowledge of light propagation through MMF can transform this low-cost optical component into a high-throughput and ultra-thin optical conduit for measuring elastic optical scattering by remote samples in a reflection regime. The demonstrated computational imaging through MMF based on round-trip measurements in a proximal spot-basis avoids the limitations in WFS and the use of fluorescent labeling, two important bottlenecks impeding practical MMF imaging. This may streamline the system design and, in combination with future advances in calibration stability of MMFs, stimulate the advent of hair-thin imaging probes that improve diagnostic performance, enhance guidance of existing interventions, and enable novel image-guided therapeutic procedures in clinical medicine.

\section{Acknowledgments}
Research in this publication was funded by the National Institute of Biomedical Imaging and Bioengineering of the National Institutes of Health, award P41 EB015903. VJP was supported by the OSA Deutsch fellowship. 

\section{Disclosures} The authors declare no conflicts of interest.

\section{Data Availability Statement} Data underlying the results presented in this paper are not publicly available at this time but may be obtained from the authors upon reasonable request.

\section{Supplemental document}
See Supplementary Materials for supporting content. 

\bibliographystyle{unsrt}

\pagebreak
\begin{center}
\textbf{\large Supplementary Materials for: \\
\bigskip
Confocal 3D reflectance imaging through multimode fibers without wavefront shaping}
\end{center}
\setcounter{section}{0}
\setcounter{equation}{0}
\setcounter{figure}{0}
\setcounter{table}{0}
\setcounter{page}{1}
\makeatletter
\renewcommand{\thesection}{S\arabic{section}}
\renewcommand{\theequation}{S\arabic{equation}}
\renewcommand{\thefigure}{S\arabic{figure}}
\renewcommand{\thetable}{S\arabic{table}} 

\section{Experimental setup}
\begin{figure}[t!] 
\centering
\includegraphics[width=5.2in]{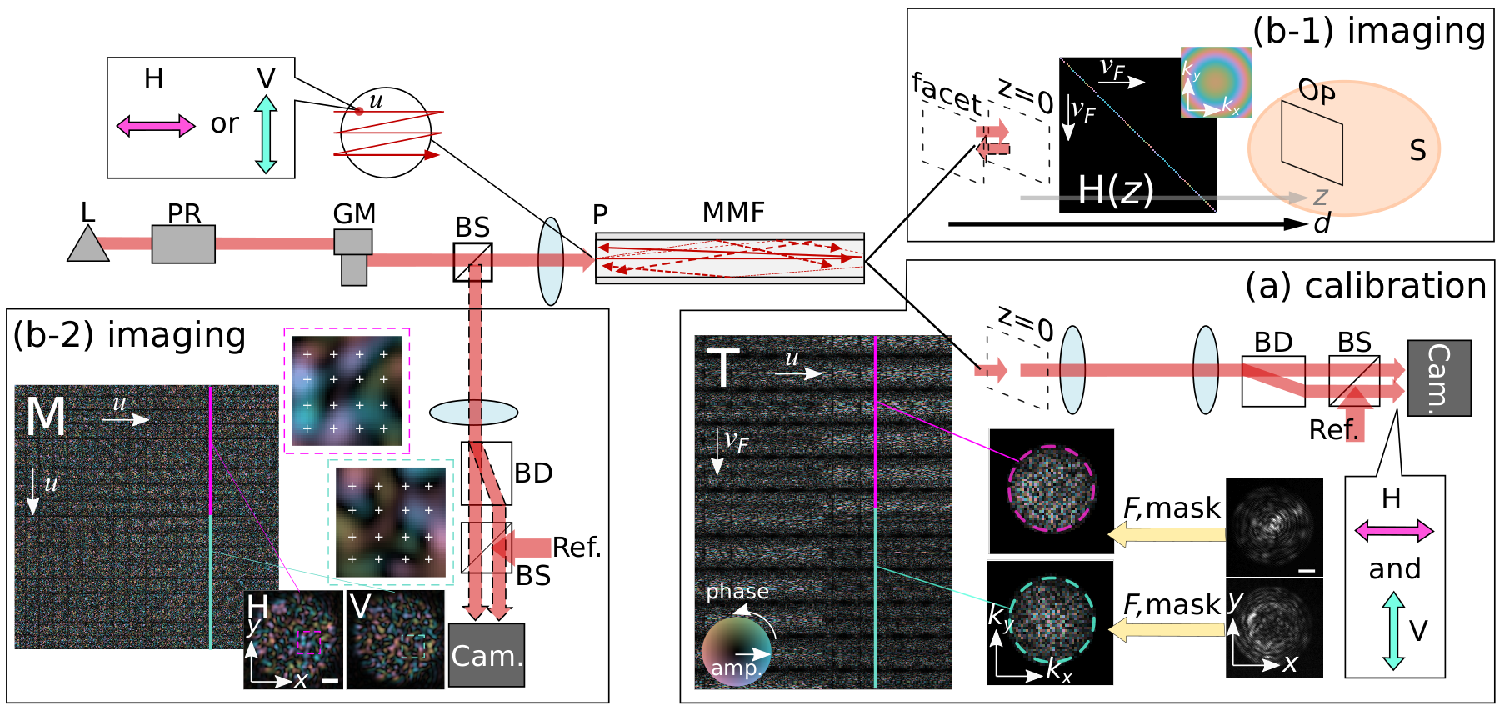}
\caption{Measurements of the MMF TMs. The fiber, although drawn as if it were straight, was in fact coiled in experiments. The red arrows correspond to light pathways, and the dashed ones indicate reflected light traveling from the sample through the MMF in the reverse direction to the proximal detection. BD: beam displacer, BS: non-polarization beam splitter, Ref.:reference wave, Cam.: camera, S: sample. A focused spot was scanned with the GM across positions indexed by $u$ distributed over the MMF proximal input facet and alternating between H and V polarizations by means of the PR. The output field was split into two orthogonal polarization states by the BD, and interfered with matching reference waves for simultaneous recording. (a) In the calibration phase, the rightmost two insets display original modulated images in ($x,y$) coordinates captured by the camera. In the Fourier domain, we isolated the demodulated complex-valued signals in momentum coordinates ($k_x,k_y$) confined to a frequency band imposed by the fiber NA and rearranged them into a column vector of \textbf{T}, as indicated by the solid vertical line, color-coded in magenta and cyan for the  H and V polarizations, respectively. The forward transmission \textbf{T} has rows and columns indexed by $\nu_F$ and $u$, respectively, and was ordered first by the spatial modes, and then by polarization states. Only a subset of \textbf{T} is shown here. The color map encodes complex values. (b-1) In the imaging phase, light backscattered from distal OPs at varying $d$ with respect to fiber facet, with free-space propagation modeled by \textbf{H}, which is a diagonal matrix in $\nu_F$ basis corresponding to a quadratic phase in Fourier domain. (b-2) The detected images were demodulated into complex-valued images of the proximal output speckle in spatial coordinates ($x,y$), then down-sampled at the positions of the input foci (shown as white markers in the dashed magenta and cyan boxes) following the same ordering as for illumination, and flattened into column vectors of the square matrix \textbf{M}. \textbf{M} thus has rows and columns indexed both by $u$. Only a subset of \textbf{M} is shown here. The scale bars in the insets are $20$ \textmu m.} 
\label{fig:expsetup}
\end{figure}

All experiments used a 1-m-long step-index MMF with $105$ \textmu m core diameter and a NA of 0.22 (FG105LCA, Thorlabs) that theoretically supports ${\sim550}$ guided modes per polarization. The fiber was coiled with a minimum radius of curvature of ${\sim50}$ mm. The monochromatic calibration matrix \textbf{T} was measured by sequentially probing the MMF input channels, while holographic detection was used for all output channels concurrently, as depicted in Fig.~\ref{fig:expsetup}(a). Each input and output channel included two orthogonal polarization states: horizontal (H) and vertical (V). To alternate the illumination polarization between H and V, a laser beam ($\lambda$ = 1550 nm and linewidth < 100 kHz) was linearly polarized and passed through a fiber-based electro-optical phase retarder (PR, Boston Applied Technologies). The laser was steered by a two-axis galvanometer scanning stage (GM, GVSM002-US, Thorlabs), and then focused by an objective lens (Plan Apo NIR Infinity Corrected, Mitutoyo) with a NA of 0.4 into a 2.5 \textmu m full-width at half maximum (FWHM) spot on the proximal facet of the MMF. The angular spectrum of the spot exceeded the NA of the MMF to ensure efficient population of all modes. The focal spot position on the proximal input side was indexed by $u$ and the speckle pattern exiting on the distal side was imaged with another identical objective lens and a tube lens ($f = 30 cm$) onto an InGaAs camera (OW1.7-VS-CL-LP-640, Raptor Photonics) with exposure time of 20 \textmu s at 120 frames per second. The object plane of the distal imaging system determined the calibration plane, which was approximately 100 \textmu m away from the distal facet. We define $d$ as the distance of the OP away from the MMF distal facet (at $d=0$). A beam displacer (BD40, Thorlabs) was used in front of the camera to spatially separate the output into H and V polarization states. An angled plane reference wave polarized at $45^{\circ}$ independently interfered with the two speckle patterns on the camera to record the speckle field amplitude and phase through off-axis holography in both detection polarization states simultaneously. Images of the two polarization states were demodulated, spatially registered, and flattened into a column vector of \textbf{T} directly in the Fourier domain, with output channels at ($k_x, k_y$) indexed by $\nu_F$. To uniformly probe all the MMF’s guided modes, transmission was recorded for an oversampled grid of input spot positions $u$, typically ${\sim}700$ points for each input polarization state, sequentially generated by driving the GM and PR. The total acquisition time was 20 seconds. The input and output spatial channels of \textbf{T} have been ordered first by spatial coordinate, then by polarization. 

In imaging experiments, as illustrated in Fig.~\ref{fig:expsetup}(b), a sample was placed in front of the MMF distal tip (b-1) and the round-trip \textbf{M} was measured from the proximal side (b-2). We again sequentially coupled light into the MMF through the same set of proximal input states. Light with a power of ~0.5 mW exited the distal facet and propagated towards the sample, where part of the light backscattered and coupled back into the same MMF. On the proximal side, we recorded the round-trip light transmission by decoupling its path from the illumination with a non-polarizing beam splitter and directing it to the same off-axis holography setup. The exposure time was set in the range 200-1000 \textmu s depending on the sample. A complete round-trip sample measurement was acquired in 20 s. To preserve the symmetry between the illumination and the detection configurations and to obtain a square matrix \textbf{M}, we sampled the recorded complex output fields at the ordered positions identical to the set of input states. The matrix \textbf{M} was then constructed with the same procedure that was used to find \textbf{T}. Numerical corrections to compensate for the physical misalignment were then applied to the output channels of the measured \textbf{M} to accurately match the input channels and recover the underlying transpose symmetry as previously described \cite{Lee:20}.

\section{Resolving axial information with numerical refocusing}
\label{sec:numerical refocusing}
For a sample with volumetric structures, under weakly scattering regime and the Born approximation, we can express the reflection matrix \textbf{R} on the calibration plane ($z=0$) as a summation of backscattering fields contributed from $N$ individual OPs at varying axial positions,
\begin{equation}
\label{eqn:R_in_rs}
\textrm{\textbf{R}} = \sum_{i=1}^{N} \textrm{\textbf{H}}^\textrm{T}(d_i) \textrm{\textbf{r}}_{d_i} \textrm{\textbf{H}}(d_i),
\end{equation}
where $\textrm{\textbf{r}}_{d_i}$, when in real-space $\nu$ representation, is a diagonal matrix with \textit{en face} reflectivity over $\textrm{i}^\textrm{th}$ OP at $z=d_i$, and \textbf{H} is a unitary TM modeling the loss-less free-space propagation from the calibration plane to the OP. Due to the unitary matrix properties,
\begin{equation}
\label{eqn:unitary_H}
\textrm{\textbf{H}}^\textrm{--1} = \textrm{\textbf{H}}^\dagger
\quad\text{and}\quad 
\textrm{\textbf{H}}^\textrm{--T} = \textrm{\textbf{H}}^\star,
\end{equation}
where the superscript $\textrm{--1}$, $\textrm{--T}$, $\dagger$, and $\star$ indicate true inverse, true inverse of transpose, Hermitian transpose, and conjugate, respectively. Note that \textbf{H} simply reduces to an identity matrix when $z=0$. According to Fresnel diffraction theory under paraxial approximation, the transfer function of a free-space propagation is a convolution kernel in real space, or a multiplicative quadratic phase term in the Fourier domain. Depending on the distance, $z$, of a selected OP, we can compute the Fourier phase term accounting for the propagation process
\begin{equation}
\label{eqn:free_space}
\textrm{$F$($k_x,k_y,z$)} = \textrm{exp( $\frac{-iz(k_x^2 + k_y^2)}{k_0}$ )},
\end{equation}
where $k_x$ and $k_y$ are the coordinates in the in-plane momentum domain. The matrix $\textrm{\textbf{H}}(z)$ in Fourier domain is then a diagonal matrix with main diagonal as $F$ at distal channels, and incorporating the matrix into \textbf{T} through left-multiplication extends the output of \textbf{T} to the OP at $z$. Plugging Eq.\ref{eqn:R_in_rs} into Eq.1 
and then into Eq.2, 
and substituting respectively
$\textrm{\textbf{T}}^\textrm{--1(tik)}$ with $\textrm{\textbf{T}}^\textrm{--1(tik)}\textrm{\textbf{H}}^\dagger(z)$ 
and 
$\textrm{\textbf{T}}^\textrm{--T(tik)}$ with $\textrm{\textbf{H}}^\star(z)\textrm{\textbf{T}}^\textrm{--T(tik)}$ 
using Eq. \ref{eqn:unitary_H}, we have a new transpose-symmetric reflection matrix \textbf{R} with input and output channels shifted to $z$
\begin{equation}
\label{eqn:R_refocus}
\tilde{\textrm{\textbf{R}}}_z = 
\textrm{\textbf{H}}^\textrm{--T}(z)\textrm{\textbf{T}}^\textrm{--T(tik)}
\textrm{\textbf{M}}
\textrm{\textbf{T}}^\textrm{--1(tik)}\textrm{\textbf{H}}^\textrm{--1}(z)
\approx 
\sum_{i=1}^{N} \textrm{\textbf{H}}^\textrm{--T}(z)\textrm{\textbf{H}}^\textrm{T}(d_i) 
\textrm{\textbf{r}}_{d_i} 
\textrm{\textbf{H}}(d_i)\textrm{\textbf{H}}^\textrm{--1}(z),
\end{equation}
which is the same as Eq. 2. If we set $z=d_j$, Eq. \ref{eqn:R_refocus} becomes
\begin{equation}
\label{eqn:R_refocus_to_d_j}
\tilde{\textrm{\textbf{R}}}_{d_j}
\approx 
\textrm{\textbf{r}}_{d_j}  +
\sum_{i\neq j}^{N} \textrm{\textbf{H}}^\textrm{--T}(d_j)\textrm{\textbf{H}}^\textrm{T}(d_i) 
\textrm{\textbf{r}}_{d_i} 
\textrm{\textbf{H}}(d_i)\textrm{\textbf{H}}^\textrm{--1}(d_j),
\end{equation}
where we isolate the in-focus and out-of-focus matrices. Assuming the out-of-focus reflective planes are separated from the in-focus plane by much more than a depth of focus, and  the total background energy is uniformly distributed over all spatial channels, we can approximate the summation of out-of-focus terms in Eq. \ref{eqn:R_refocus_to_d_j} as a complex matrix with random phases but a constant amplitude. Collecting the on-diagonal elements of $\tilde{\textrm{\textbf{R}}}_{d_j}$ hence leads to signal predominance by the \textit{en face} reflectivity at $z=d_j$ and suppression of out-of-focus signals, or background rejection. In short, after measuring \textbf{M}, by obtaining $\tilde{\textrm{\textbf{R}}}$ at intended axial position following Eq. \ref{eqn:R_refocus} and then applying Eq.3 
, we can digitally shift to the $\textrm{j}^\textrm{th}$ OP and image the \textit{en face} reflectivity at $z=d_j$ without repeated measurements. 

\section{Digital resampling of image physical and digital dimensions}
\label{sec:resampling}
The light transport through a MMF and interaction with a distal sample can be well modeled with measured TMs, which contain full complex propagation information of wave-vectors within the NA of the MMF. While the experimental \textbf{T} has output channels stored in Fourier domain, with an one-time measured \textbf{M} in the imaging phase, arbitrary resampling of 2D image dimensions and also digital adjustment of image size on any selected OPs can be readily configured based on Fourier relations. This offers flexible trade-off between image processing speed and accuracy in a pragmatic circumstance: a lower resampling rate or smaller physical dimension reduces the computational burden, which is suitable for a faster image preview, whereas a higher resampling rate produces a detailed and smooth image at the expense of longer processing duration. Here, we quantify the trade-off by timing the image processing on a personal computer with a 3.4 GHz Intel Core i7 CPU and 16 GB RAM using MATLAB. 

For an arbitrary setting of image physical and digital dimensions, we upsampled the output spatial channels of \textbf{T} in the Fourier domain by interpolation, and pre-computed an inverse discrete Fourier Transform (iDFT) matrix for converting the distal channels to resampled real-space coordinates during the 2D real-space image reconstruction. We focus on the upsampling that corresponds to a valid augmentation to the initial pupil size on the calibration plane (${\sim}105$ \textmu m in diameter). Note that the interpolation of \textbf{T} output channels and the calculation of iDFT matrices are performed prior to actual image formation processes. The necessary computation of images on an OP involves application of phase terms to \textbf{T} outputs for intended numerical refocusing, distal spatial channels conversion into real-space coordinates with the prepared iDFT matrix, left and right multiplication of \textbf{M} with regularized inversion of extended backward and forward TMs following Eq. \ref{eqn:R_refocus} to retrieve $\tilde{\textrm{\textbf{{R}}}}$, and reshaping back to a 2D image using Eq.3. 

\begin{figure}[t!] 
\centering
\includegraphics[width=5in]{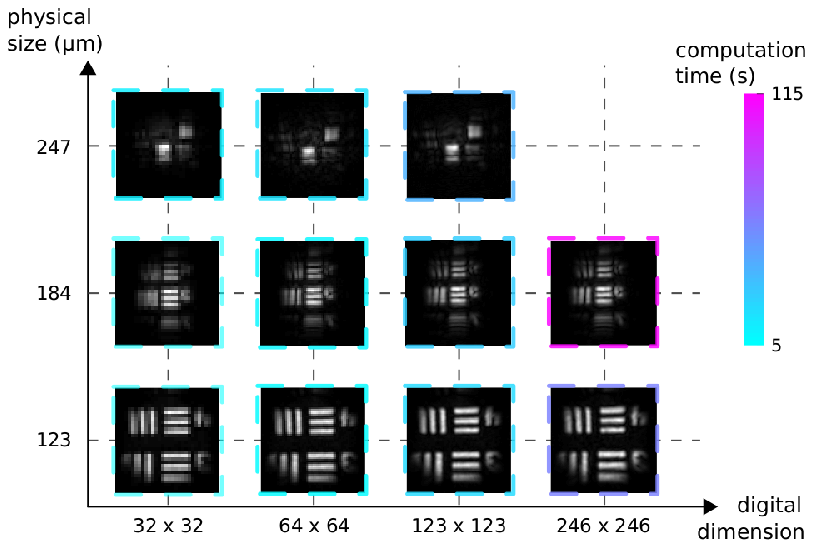}
\caption{Computation time and quality of confocal images as a function of physical and digital image size. We only show intensity images in a single polarization state since the sample is binary and isotropic. The time in second is color coded.}
\label{fig:resampling}
\end{figure}

In the experiment, the initial \textbf{T} had output channels accounting for $247 \times 247$ square area of camera recording pixels conjugating a physical size of $123 \times 123$ \textmu m$^2$, and the resolution chart as sample was placed on an OP at $d=10,\:600,\:1200$ \textmu m away from the facet. For each imaging setting, we timed only the necessary computation. As shown in Fig. \ref{fig:resampling}, the computation for co-polarization 2D confocal images at $d=10$ \textmu m with original dimensions and size takes ${\sim}58.2$ sec. To reduce computation complexity and complete image formation in a shorter time, we can down-sample the image dimensions to $32 \times 32$ in the same physical extent, resulting in pixelated images on OPs at $d=10$ \textmu m calculated within ${\sim}5.2$ sec. For images on an OP at $d=1200$ \textmu m distant from the distal MMF facet, illuminating light diverges, and a larger configured image physical dimension is needed to avoid image clipping. For instance, the computation time of $32 \times 32$ confocal intensity images covering $247 \times 247$ \textmu m$^2$ on the OP at $d=1200$ \textmu m is ${\sim11.2}$ sec. Table \ref{table:computation time} summarizes the computation time of individual settings.

\begin{table}[h!]
\centering
\begin{tabular}{ |c|c|c|c|c|  }
 \hline
 \multicolumn{5}{|c|}{computation time (sec)} \\
 \hline
 physical size (\textmu m)$\backslash$digital dimension &32 &64 &123 &247 \\
 \hline
 123    &5.2  &7.9  &17.5 & 58.2\\
 \hline
 184    &7.4  &13.7 &28.5 & 114.81\\
 \hline
 247    &11.2 &20.3 &39.9 & \\
 \hline
\end{tabular}
\caption{Computation time of confocal images considering different image configuration settings.}
\label{table:computation time}
\end{table}

\section{3D resolution and field of view}
\label{sec:resolution and FOV theory}
\begin{figure}[t!] 
\centering
\includegraphics[width=5in]{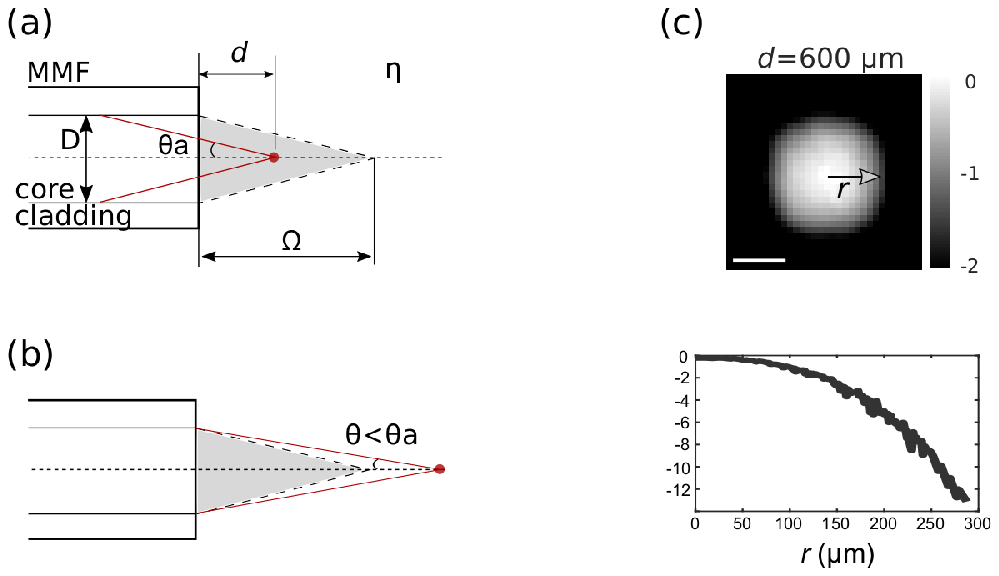}
\caption{Illustration of effective NA for an on-axis point object (red dot) on an OP (a) within and (b) beyond the MMF focal length. The dotted line indicates optical axis. (c) The simulated beam divergence at $d=600$ \textmu m with experimental \textbf{T}, and the circular blob diameter associates with the imaging FOV. The plotted radius-wise mean intensity in logarithmic scale with defined threshold determines the expected FOV on the OP at $d$. The scale bar is $100$ \textmu m.} 
\label{fig:effective NA}
\end{figure}

To calculate the theoretical lateral and axial resolution, we need to first compute the effective NA, $\mathit{NA_{eff}}$, specific to an OP at an axial position. While the effective NA may also be dependent of the lateral displacement from the optical axis, we only consider an on-axis point object on the OP for convenience. Given the object at a distance $d$ away from the MMF facet, the effective NA can be calculated from the maximal angle formed with the point as the vertex and marginal rays within the MMF acceptance angle as sides, as illustrated in Fig. \ref{fig:effective NA}(a) and (b). When $d$ is within the focal length of the MMF, $\Omega\ {\sim}\ \eta D/2\mathit{NA}$, a full NA can be obtained, which is determined during MMF fabrication. Here, $\eta$ is the medium refractive index, and $\theta_a$ is the fiber acceptance angle. Once $d$ is larger than this range, only a partial NA can be achieved due to the limited MMF diameter. The value of effective NA is summarized as
\begin{equation}
\label{eqn:effective NA}
\mathit{NA_{eff}}=\begin{cases}
    \mathit{NA}, & \text{if $d<\frac{\eta D}{2\mathit{NA}}$}.\\
    \sim\frac{D}{2d}, & \text{otherwise}. 
   \end{cases}
\end{equation}
Given the effective NA, we can then compute the expected lateral and axial resolution as in confocal microscopy \cite{Spring}
\begin{subequations}
\begin{align} \label{eqn:resolution}
\delta x = \frac{0.4 \:\lambda}{\mathit{NA_{eff}}} \\
\delta z = \frac{1.4 \:\eta \:\lambda}{\mathit{NA^2_{eff}}},
\end{align}
\end{subequations}
where we see that the axial resolution has a strong dependence on the system NA.

With the circular symmetry of fiber core shape, we can define the FOV on an OP as the diameter of a circular area with circumference from furthest off-axis points having normalized confocally detected intensity dropped below $1$\% threshold. Using the measured \textbf{T} of the MMF, we can free-space propagate each output light field per input to an OP and incoherently sum all output light intensity over each input realization. This results in a circular blob on the OP indicating the average illumination power at each spatial channel. Taking the spatial-channel-wise intensity square of the blob informs confocally detectable power, as shown in Fig. \ref{fig:effective NA}(c), where the OP is $600$ \textmu m away from the MMF distal facet. The low light coupling efficiency at FOV peripheral causes the vignetting effect on reconstructed images, and the quantified FOV has $\varnothing{\sim}260$ \textmu m by applying the threshold to plotted radius-wise mean intensity.

\section{Reconstructing confocal images from partial TM measurement}
\label{sec:imaging from partial TM}
\begin{figure}[h!] 
\centering
\includegraphics[width=5.2in]{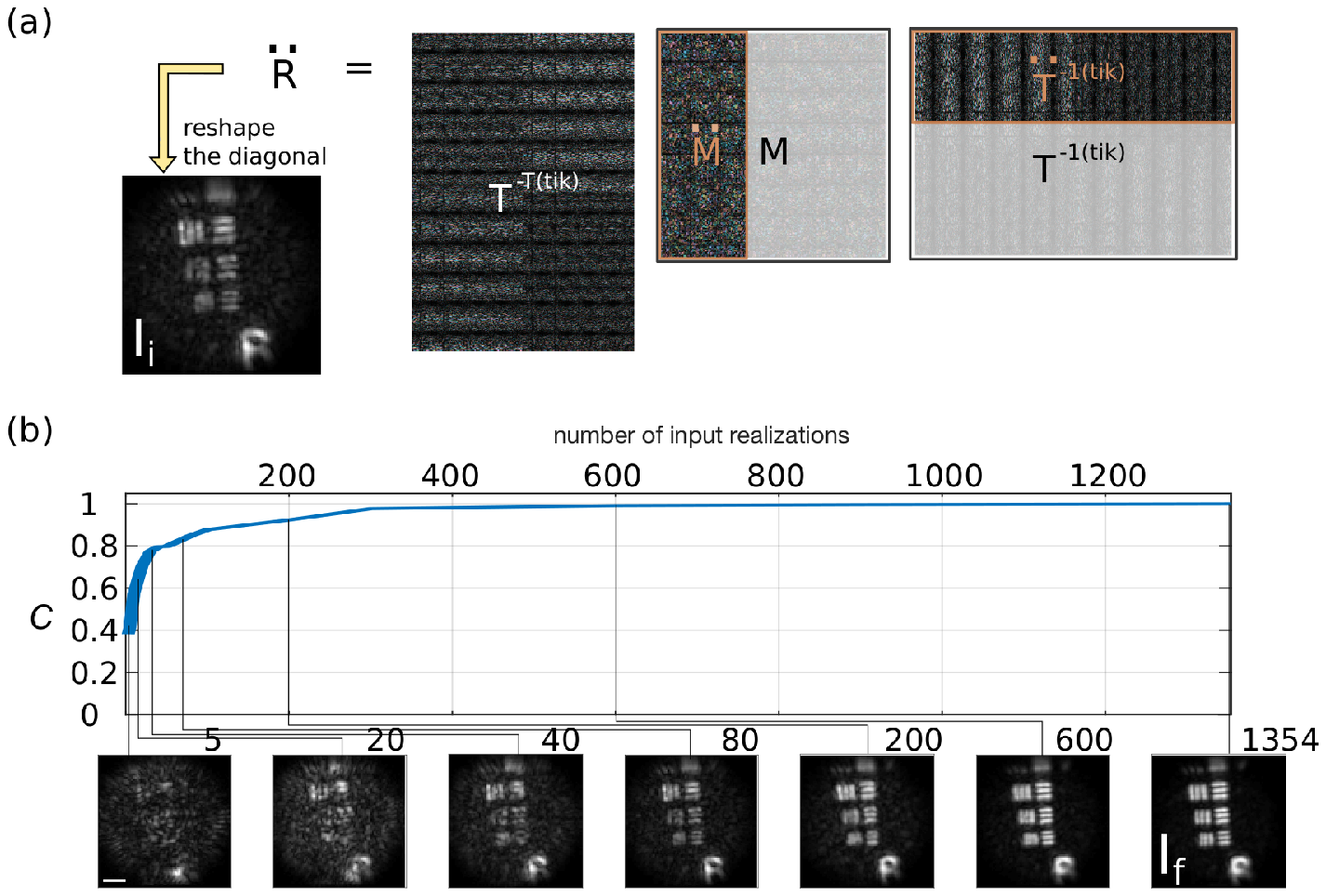}
\caption{Confocal intensity image reconstruction with partial measurement (a) Illustration of computational image reconstruction with the MMF forward TM, \textbf{T}, and a partially measured round-trip reflection matrix, \textbf{\"{M}} (b) Reconstruction results and image correlation with increasing round-trip measurement completeness. The scale bar is $20$ \textmu m.} 
\label{fig:confocal with partial TM}
\end{figure}

In the imaging phase, we can reconstruct confocal images from the round-trip measurement of \textbf{M} by obtaining the reflection matrix, $\tilde{\textrm{\textbf{R}}}$, processing its elements, and reshaping into 2D real-space coordinates at a selected OP. While the measurement of a full \textbf{M} by sequentially coupling light into all MMF proximal channels delivers maximal information of the distal sample bounded by the MMF throughput, intermediate confocal images for preview can also be reconstructed from a round-trip measurement with partial set of input realizations, \textbf{\"{M}}, which is a subset of \textbf{M} containing constituent column vectors, leading to a rectangular matrix. As illustrated in Figure \ref{fig:confocal with partial TM}(a), with \textbf{\"{M}}, we can reconstruct an speckled image on an OP from a computed reflection matrix, \textbf{\"{R}}, by respectively left and right multiplying \textbf{\"{M}} with full $\textrm{\textbf{T}}^{\textrm{-T(tik)}}$ and $\textrm{\textbf{\"{T}}}^{\textrm{-1(tik)}}$, which is the regularized inverse of a subset of \textbf{T} with constituent column vectors at input channels corresponding to \textbf{\"{M}}. Physically speaking, the image derived from the partial measurement corresponds to the distal sample under statistically non-uniform illumination. Using confocal intensity images \textbf{I} for demonstration here, we define the completeness of an intermediate image as the normalized intensity correlation, $C$, with the final image reconstructed from full \textbf{M} measurement, 
\begin{equation}
\label{eqn:image correlation}
C = \frac{\sum_{x,y}\textrm{\textbf{I}}_i(x,y)\textrm{\textbf{I}}_f(x,y)}{\sum_{x,y}\textrm{\textbf{I}}_i(x,y)\sum_{x,y}\textrm{\textbf{I}}_f(x,y)},
\end{equation}
where $\textrm{\textbf{I}}_i$ and $\textrm{\textbf{I}}_f$ are intermediate and final images, respectively. The completeness arrives at $C=1$ when $\textrm{\textbf{I}}_i=\textrm{\textbf{I}}_f$. Figure \ref{fig:confocal with partial TM}(b) shows examples of intermediate images with their quantified completeness. Here, the sample is a resolution chart, and the full \textbf{M} is a $1354$-by-$1354$ square matrix. From the plot, we can see that the completeness quickly improves with the number of input realizations and achieves $90$\% with ${\sim}200$ input realizations, which is only ${\sim}15\%$ of the total number of realizations. The intermediate images start from speckled pattern and evolve to clean and high-contrast final confocal intensity image.

\section{3D confocal images with various contrasts}
\label{sec:3D imaging}
\begin{figure}[htbp!] 
\centering
\includegraphics[width=3.5in]{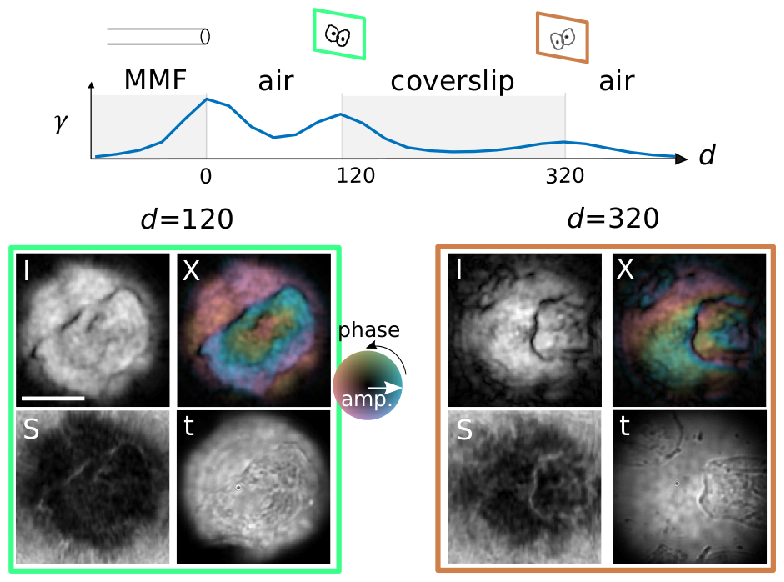}
\caption{Label-free 3D computational imaging through the MMF with multiple contrasts. The sample included two layers of buccal epithelial cells deposited on both surfaces of a glass coverslip in air. The $d$ position indicates physical distance and is in the unit of \textmu m. The scale bars are $50$ \textmu m.}
\label{fig:3D_bead_cell_imaging}
\end{figure}

To demonstrate 3D imaging of biological samples through the MMF based on numerical refocusing, a sample with multiple layers was prepared following similar volumetric reconstruction experiments performed by others \cite{Tsvirkun:16,Shineaaw19,choi2020fourier}. A proximal reflectance measurement of \textbf{M} through the MMF included reflectance from multiple layers of a sample, shown in Fig. \ref{fig:3D_bead_cell_imaging}(top), including buccal epithelial cells deposited on both surfaces of a glass coverslip with thickness of ${\sim}200$ \textmu m, placed at $d=120$ \textmu m in air. From this single \textbf{M}, 3D volumetric imaging was computed by numerical refocusing and image reconstruction. The depth-dependent $\gamma$ plot was consistent with the physical location of each reflective interface, considering the refractive indices of each layer (1.44 in glass). High-resolution confocal images with intensity, phase, and scattering contrasts were computed at the two individual coverslip surfaces ($d=120$ and $320$ \textmu m). Both planes exhibited contrast from cell samples in all images, consistent with the transmission ground truth, and with high contrast, indicating the confocal gating efficacy. Note that in complex samples, because optical phase accumulates as light is reflected from further into the sample, the phase of shallower cells is overlaid on deeper-lying cells.

\bibliographystyle{unsrt}

\begin{thebibliography}{10}

\bibitem{Keiser:14}
Gerd Keiser, Fei Xiong, Ying Cui, and Perry~Ping Shum.
\newblock {Review of diverse optical fibers used in biomedical research and
  clinical practice}.
\newblock {\em Journal of Biomedical Optics}, 19(8):1 -- 29, 2014.

\bibitem{Loterie:17}
Damien Loterie, Demetri Psaltis, and Christophe Moser.
\newblock Bend translation in multimode fiber imaging.
\newblock {\em Opt. Express}, 25(6):6263--6273, Mar 2017.

\bibitem{PhysRevX.9.041050}
George S.~D. Gordon, Milana Gataric, Alberto Gil C.~P. Ramos, Ralf Mouthaan,
  Calum Williams, Jonghee Yoon, Timothy~D. Wilkinson, and Sarah~E. Bohndiek.
\newblock Characterizing optical fiber transmission matrices using metasurface
  reflector stacks for lensless imaging without distal access.
\newblock {\em Phys. Rev. X}, 9:041050, Dec 2019.

\bibitem{matthes2020learning}
Maxime~W. MatthÃ¨s, Yaron Bromberg, Julien de~Rosny, and SÃ©bastien~M. Popoff.
\newblock Learning and avoiding disorder in multimode fibers, 2020.

\bibitem{li2020guidestar}
Shuhui Li, Simon A.~R. Horsley, Tomas Tyc, Tomas Cizmar, and David~B. Phillips.
\newblock Guide-star assisted imaging through multimode optical fibres, 2020.

\bibitem{PhysRevLett.109.203901}
Youngwoon Choi, Changhyeong Yoon, Moonseok Kim, Taeseok~Daniel Yang,
  Christopher Fang-Yen, Ramachandra~R. Dasari, Kyoung~Jin Lee, and Wonshik
  Choi.
\newblock Scanner-free and wide-field endoscopic imaging by using a single
  multimode optical fiber.
\newblock {\em Phys. Rev. Lett.}, 109:203901, Nov 2012.

\bibitem{Turtaev2018}
Sergey Turtaev, Ivo~T. Leite, Tristan Altwegg-Boussac, Janelle M.~P. Pakan,
  Nathalie~L. Rochefort, and Tom{\'{a}}{\v{s}} {\v{C}}i{\v{z}}m{\'{a}}r.
\newblock {High-fidelity multimode fibre-based endoscopy for deep brain in vivo
  imaging}.
\newblock {\em Light: Science {\&} Applications}, 7(1):92, dec 2018.

\bibitem{Vasquez-Lopez:18}
Sebastian~A. Vasquez-Lopez, Rapha{\"e}l Turcotte, Vadim Koren, Martin
  Pl{\"o}schner, Zahid Padamsey, Martin~J. Booth, Tom{\'a}{\v{s}}
  {\v{C}}i{\v{z}}m{\'a}r, and Nigel~J. Emptage.
\newblock Subcellular spatial resolution achieved for deep-brain imaging in
  vivo using a minimally invasive multimode fiber.
\newblock {\em Light: Science {\&} Applications}, 7(1):110, Dec 2018.

\bibitem{Leite:21}
Ivo~T. Leite, Sergey Turtaev, Dirk~E. Boonzajer~Flaes, and Tom{\'a}{\v{s}}
  {\v{C}}i{\v{z}}m{\'a}r.
\newblock Observing distant objects with a multimode fiber-based holographic
  endoscope.
\newblock {\em APL Photonics}, 6(3):036112, Mar 2021.

\bibitem{Papadopoulos:12}
Ioannis~N. Papadopoulos, Salma Farahi, Christophe Moser, and Demetri Psaltis.
\newblock Focusing and scanning light through a multimode optical fiber using
  digital phase conjugation.
\newblock {\em Opt. Express}, 20(10):10583--10590, May 2012.

\bibitem{Cizmar2012}
Tom{\'{a}}{\v{s}} {\v{C}}i{\v{z}}m{\'{a}}r and Kishan Dholakia.
\newblock {Exploiting multimode waveguides for pure fibre-based imaging}.
\newblock {\em Nature Communications}, 3, 2012.

\bibitem{Loterie:15}
Damien Loterie, Salma Farahi, Ioannis Papadopoulos, Alexandre Goy, Demetri
  Psaltis, and Christophe Moser.
\newblock Digital confocal microscopy through a multimode fiber.
\newblock {\em Opt. Express}, 23(18):23845--23858, Sep 2015.

\bibitem{Ploschner2015}
Martin Pl{\"{o}}schner, Tom{\'{a}}{\v{s}} Tyc, and Tom{\'{a}}{\v{s}}
  {\v{C}}i{\v{z}}m{\'{a}}r.
\newblock {Seeing through chaos in multimode fibres}.
\newblock {\em Nature Photonics}, 9(8):529--535, 2015.

\bibitem{Gusachenko:17}
Ivan Gusachenko, Mingzhou Chen, and Kishan Dholakia.
\newblock Raman imaging through a single multimode fibre.
\newblock {\em Opt. Express}, 25(12):13782--13798, Jun 2017.

\bibitem{Ohayon:18}
Shay Ohayon, Antonio Caravaca-Aguirre, Rafael Piestun, and James~J. DiCarlo.
\newblock Minimally invasive multimode optical fiber microendoscope for deep
  brain fluorescence imaging.
\newblock {\em Biomed. Opt. Express}, 9(4):1492--1509, Apr 2018.

\bibitem{Caravaca-Aguirre:19}
Antonio~M. Caravaca-Aguirre, Sakshi Singh, Simon Labouesse, Michael~V. Baratta,
  Rafael Piestun, and Emmanuel Bossy.
\newblock Hybrid photoacoustic-fluorescence microendoscopy through a multimode
  fiber using speckle illumination.
\newblock {\em APL Photonics}, 4(9):096103, 2019.

\bibitem{Amitonova2018}
Lyubov~V. Amitonova and Johannes~F. de~Boer.
\newblock {Compressive imaging through a multimode fiber}.
\newblock {\em Optics Letters}, 43(21), 2018.

\bibitem{Choudhury2020}
Debaditya Choudhury, Duncan~K. McNicholl, Audrey Repetti, Itandehui
  Gris-S{\'a}nchez, Shuhui Li, David~B. Phillips, Graeme Whyte, Tim~A. Birks,
  Yves Wiaux, and Robert~R. Thomson.
\newblock Computational optical imaging with a photonic lantern.
\newblock {\em Nature Communications}, 11(1):5217, Oct 2020.

\bibitem{Borhani:18}
Navid Borhani, Eirini Kakkava, Christophe Moser, and Demetri Psaltis.
\newblock Learning to see through multimode fibers.
\newblock {\em Optica}, 5(8):960--966, Aug 2018.

\bibitem{Caramazza2019}
Piergiorgio Caramazza, Ois{\'i}n Moran, Roderick Murray-Smith, and Daniele
  Faccio.
\newblock Transmission of natural scene images through a multimode fibre.
\newblock {\em Nature Communications}, 10(1):2029, May 2019.

\bibitem{Stellinga:19}
Daan Stellinga, David~B. Phillips, Adam Selyem, Sergey Turtaev, Tom\'{a}\v{s}
  \v{C}i\v{z}m\'{a}r, and Miles~J. Padgett.
\newblock Time of flight based 3d imaging through multimode optical fibres.
\newblock In {\em Frontiers in Optics $+$ Laser Science APS/DLS}, page FW6F.4.
  Optical Society of America, 2019.

\bibitem{Tzang2019}
Omer Tzang, Eyal Niv, Sakshi Singh, Simon Labouesse, Greg Myatt, and Rafael
  Piestun.
\newblock Wavefront shaping in complex media with a 350{\thinspace}khz
  modulator via a 1d-to-2d transform.
\newblock {\em Nature Photonics}, 13(11):788--793, Nov 2019.

\bibitem{Lee:20}
Szu-Yu Lee, Vicente~J. Parot, Brett~E. Bouma, and Martin Villiger.
\newblock Reciprocity-induced symmetry in the round-trip transmission through
  complex systems.
\newblock {\em APL Photonics}, 5(10):106104, 2020.

\bibitem{Hansen00thel-curve}
P.~C. Hansen.
\newblock The l-curve and its use in the numerical treatment of inverse
  problems.
\newblock In {\em in Computational Inverse Problems in Electrocardiology, ed.
  P. Johnston, Advances in Computational Bioengineering}, pages 119--142. WIT
  Press, 2000.

\bibitem{PhysRevLett.107.023902}
Youngwoon Choi, Taeseok~Daniel Yang, Christopher Fang-Yen, Pilsung Kang,
  Kyoung~Jin Lee, Ramachandra~R. Dasari, Michael~S. Feld, and Wonshik Choi.
\newblock Overcoming the diffraction limit using multiple light scattering in a
  highly disordered medium.
\newblock {\em Phys. Rev. Lett.}, 107:023902, Jul 2011.

\bibitem{Scoles:13}
Drew Scoles, Yusufu~N. Sulai, and Alfredo Dubra.
\newblock In vivo dark-field imaging of the retinal pigment epithelium cell
  mosaic.
\newblock {\em Biomed. Opt. Express}, 4(9):1710--1723, Sep 2013.

\bibitem{Yoon2017}
Changhyeong Yoon, Munkyu Kang, Jin~H. Hong, Taeseok~D. Yang, Jingchao Xing,
  Hongki Yoo, Youngwoon Choi, and Wonshik Choi.
\newblock Removal of back-reflection noise at ultrathin imaging probes by the
  single-core illumination and wide-field detection.
\newblock {\em Scientific Reports}, 7(1):6524, Jul 2017.

\bibitem{Liu2020}
Jian Liu, Jing Liu, Chenguang Liu, and Yuhang Wang.
\newblock {3D dark-field confocal microscopy for subsurface defects detection}.
\newblock {\em Optics Letters}, 45(3):660, 2020.

\bibitem{Morales-Delgado:15}
Edgar~E. Morales-Delgado, Demetri Psaltis, and Christophe Moser.
\newblock Two-photon imaging through a multimode fiber.
\newblock {\em Opt. Express}, 23(25):32158--32170, Dec 2015.

\bibitem{OH2013760}
Gyungseok Oh, Euiheon Chung, and Seok~H. Yun.
\newblock Optical fibers for high-resolution in vivo microendoscopic
  fluorescence imaging.
\newblock {\em Optical Fiber Technology}, 19(6, Part B):760--771, 2013.
\newblock Optical Fiber Sensors.

\bibitem{Gu2014}
M.~GU, H.~BAO, and H.~KANG.
\newblock Fibre-optical microendoscopy.
\newblock {\em Journal of Microscopy}, 254(1):13--18, 2014.

\bibitem{Boniface:17}
Antoine Boniface, Mickael Mounaix, Baptiste Blochet, Rafael Piestun, and
  Sylvain Gigan.
\newblock Transmission-matrix-based point-spread-function engineering through a
  complex medium.
\newblock {\em Optica}, 4(1):54--59, Jan 2017.

\bibitem{Caravaca-Aguirre:13}
Antonio~M. Caravaca-Aguirre, Eyal Niv, Donald~B. Conkey, and Rafael Piestun.
\newblock Real-time resilient focusing through a bending multimode fiber.
\newblock {\em Opt. Express}, 21(10):12881--12887, May 2013.

\bibitem{Gu:15}
Ruo~Yu Gu, Reza~Nasiri Mahalati, and Joseph~M. Kahn.
\newblock Design of flexible multi-mode fiber endoscope.
\newblock {\em Opt. Express}, 23(21):26905--26918, Oct 2015.

\bibitem{Li2021}
Shuhui Li, Charles Saunders, Daniel~J. Lum, John Murray-Bruce, Vivek~K. Goyal,
  Tom{\'a}{\v{s}} {\v{C}}i{\v{z}}m{\'a}r, and David~B. Phillips.
\newblock Compressively sampling the optical transmission matrix of a multimode
  fibre.
\newblock {\em Light: Science {\&} Applications}, 10(1):88, Apr 2021.

\end{thebibliography}

\begin{thebibliography}{1}

\bibitem{Spring}
T.~J.~F. Kenneth R.~Spring and M.~W. Davidson.
\newblock Resolution and contrast in confocal microscopy
\newblock Olympus - Life Science Solution.

\bibitem{Tsvirkun:16}
V.~Tsvirkun, S.~Sivankutty, G.~Bouwmans, O.~Katz, E.~R. Andresen, and
  H.~Rigneault. 
\newblock Widefield lensless endoscopy with a multicore fiber
\newblock Opt. Lett. \textbf{41}, 4771--4774 (2016).

\bibitem{Shineaaw19}
J.~Shin, D.~N. Tran, J.~R. Stroud, S.~Chin, T.~D. Tran, and M.~A. Foster.
\newblock A minimally invasive lens-free computational microendoscope.
\newblock Science Advances \textbf{5} (2019).

\bibitem{choi2020fourier}
W.~Choi, M.~Kang, J.~H. Hong, O.~Katz, Y.~Choi, and W.~Choi.
\newblock Fourier holographic endoscopy for label-free imaging through a narrow and curved passage. 
\newblock arXiv (2020).

\end{thebibliography}

\end{document}